\documentclass[12pt]{amsart}
\usepackage{amsmath,amssymb,float,amsthm}
\usepackage{graphicx}
\usepackage{bm}

\usepackage{hyperref}
\usepackage[margin = 1.15in]{geometry}

\newcommand{\ubar}{\overline{u}}
\renewcommand{\hat}{\widehat} 
\newcommand{\cc}{\mathrm{c.c.}}

\newtheorem{assumption}{Assumption}
\newtheorem{lemma}{Lemma}
\newtheorem{proposition}{Proposition}
\newtheorem{corollary}{Corollary}

\usepackage{xcolor}
\definecolor{forest}{rgb}{0.13, 0.55, 0.13}

\title[Whitham modulation theory for generalized Whitham equations]{Whitham modulation theory for generalized Whitham equations and a general criterion for modulational instability}

\author[A. L. Binswanger]{Adam L. Binswanger$^{1}$}
\email{abinswanger@ucmerced.edu}
\author[M. A. Hoefer]{Mark A. Hoefer$^{2}$}
\author[B. Ilan]{Boaz Ilan$^{1}$}
\author[P. Sprenger]{Patrick Sprenger$^{2^*,3}$}
\address{$^{1}$School of Natural Sciences, University of California Merced, Merced, USA}%
\address{$^{2}$ Department of Applied Mathematics, University  of Colorado Boulder, Boulder, USA} 
\address{$^{3}$ Department of Mathematics, North Carolina State
  University, Raleigh, USA}
\address{$^{*}$ Former affiliation} 

\date{\today}

\begin{document}

\maketitle


\begin{abstract}
The Whitham equation was proposed as a model for surface water waves that combines the quadratic flux nonlinearity $f(u) = \tfrac{1}{2}u^2$ of the Korteweg-de Vries equation and the full linear dispersion relation $\Omega(k) = \sqrt{k\tanh k}$ of uni-directional gravity water waves in suitably scaled variables. This paper proposes and analyzes a generalization of Whitham's model to unidirectional nonlinear wave equations consisting of a general nonlinear flux function $f(u)$ and a general linear dispersion relation $\Omega(k)$. Assuming the existence of periodic traveling wave solutions to this generalized Whitham equation, their slow modulations are studied in the context of Whitham modulation theory. A multiple scales calculation yields the modulation equations, a system of three conservation laws that describe the slow evolution of the periodic traveling wave's wavenumber, amplitude, and mean. In the weakly nonlinear limit, explicit, simple criteria in terms of general $f(u)$ and $\Omega(k)$ establishing the strict hyperbolicity and genuine nonlinearity of the modulation equations are determined. This result is interpreted as a generalized Lighthill-Whitham criterion for modulational instability.
\end{abstract}


\section{Introduction}

Scalar dispersive hydrodynamic equations model nonlinear wave motion in systems where dissipation is negligible with respect to wave dispersion. Many nonlinear wave models are derived using a multiple scale procedure in the small amplitude, long wavelength regime \cite{ablowitz_nonlinear_2011}. The canonical model equation that describes the unidirectional propagation of weakly nonlinear long waves is the Korteweg-de Vries (KdV) equation 
\begin{align}\label{eq:kdv}
u_t + \left(\frac{1}{2}u^2\right)_x + u_{xxx} = 0,
\end{align}
which arises universally in systems exhibiting weak quadratic nonlinearity and long-wave, third order dispersion \cite{whitham_linear_1974}.  Equation \eqref{eq:kdv} can be derived, for example, as a model of the free surface displacement $u(x,t)$ of an incompressible fluid with motion in one spatial dimension. In this scenario, the KdV equation \eqref{eq:kdv} provides an accurate description of free surface dynamics for some phenomena--for instance the evolution of broad disturbances--but fails to capture short wave phenomena such as sharp profiles, e.g. peaking waves. In a 1967 paper, Whitham proposed an alternative model equation consisting of the same quadratic nonlinearity as the KdV equation \eqref{eq:kdv} while capturing the full linear dispersion of waves moving in the positive $x$-direction in order to model wave peaking \cite{whitham_variational_1967}. The so-called Whitham equation for surface water waves is given in nondimensional coordinates by  
\begin{align}\label{eq:whitham_quadNL}
u_t + \left(\frac{1}{2} u^2\right)_x + \mathcal{K}*u_x = 0, 
\end{align}
where $\mathcal{K}*u_x$ is defined as a Fourier multiplier with the symbol 
\begin{align}\label{eq:whitham_ww_ker}
\widehat{\mathcal{K}*g(x)} = \sqrt{\frac{\tanh q}{q}} \hat{g}(q), 
\end{align}
where $q$ is the Fourier wavenumber coordinate and $\hat{g}(q)$ is the Fourier transform of $g(x)$ defined by the pair 
\begin{align}\label{eq:fourier_trans_def}
  \begin{split}
    g(x) &= \frac{1}{2\pi} \int_{\mathbb{R}} \widehat{g}(q) e^{iqx}\,
    \mathrm{d}q, \\
    \widehat{g}(q) &= \int_{\mathbb{R}} g(x) e^{-iqx}\,
    \mathrm{d}x,
  \end{split}
\end{align}
for $g(x) \in L^2(\mathbb{R})$. Thus $\mathcal{K}$ is a convolution kernel whose Fourier transform is the linear phase velocity for unidirectional water waves $\sqrt{\tanh q/q}$. The Whitham equation \eqref{eq:whitham_ww_ker} can be derived via an asymptotic expansion of the Hamiltonian formulation of the Euler equations with free surface boundary conditions \cite{moldabayev_whitham_2015,dinvay_whitham_2019}. The existence and numerical computation of smooth periodic, traveling wave solutions was established in \cite{ehrnstrom_traveling_2009}. Subsequently,  the existence of limiting periodic solutions with a cusp, i.e. peaked waves, has been established \cite{ehrnstrom_global_2013}. Periodic solutions were shown to be unstable with respect to long wavelength perturbations in sufficiently deep water \cite{sanford_stability_2014}.  Solitary wave solutions are also known to exist \cite{ehrnstrom_existence_2012}. From a modeling perspective, Eq. \eqref{eq:whitham_ww_ker} outperforms the KdV equation in its approximation of the free surface dynamics of a water wave. This is shown both in numerical simulation of the Euler equations \cite{moldabayev_whitham_2015} and in experimental wave tanks \cite{trillo_observation_2016,carter_bidirectional_2018}.  The success of the Whitham equation \eqref{eq:whitham} as a model for surface water waves motivates consideration of full-dispersion models in the context of other physical systems. For example, models similar to equation \eqref{eq:whitham_quadNL} with a modified convolution operator have been used to model water waves underneath an elastic ice sheet \cite{dinvay_whitham_2019} and internal waves at a two-fluid interface, one with infinite depth \cite{joseph_solitary_1977}.

In this manuscript, we propose a generalization of \eqref{eq:whitham_quadNL} where the quadratic nonlinear term is replaced by a general flux and the linear dispersion is similarly determined by a convolution. The generalized Whitham equation takes the form 
\begin{equation}
\label{eq:whitham}
    u_t + f(u)_{x} + \mathcal{K}*u_{x} = 0,
\end{equation}
where the convolution $\mathcal{K}*g(x)$ is defined by the Fourier multiplier
\begin{align*}
\widehat{\mathcal{K}*g(x)} = \frac{\Omega(q)}{q} \hat{g}(q).
\end{align*}
Here, $f(u)$ and $\Omega(k)$ are the problem dependent nonlinear hydrodynamic flux and linear dispersion relation, respectively. The dispersion relation of small amplitude, sinusoidal waves on a nonzero background mean, $\ubar$, depend on the nonlinear flux and linear dispersion according to  
\begin{align}
\omega(\ubar,k)  = f'(\ubar)k + \Omega(k),
\end{align} 
where $k$ is the wavenumber. It should be noted that a more complex dependence on the wave mean of the linear dispersion, e.g. in models with nonlinear dispersive terms, is not included in this generalized Whitham equation \eqref{eq:whitham} (see ref. \cite{lannes_water_2013} for a discussion of other full dispersion models for water waves). However, the class of generalized Whitham equations \eqref{eq:whitham} is quite general.  Setting $f(u) = \frac{1}{2}u^2$ and $\Omega(k) = -k^3$ recovers the KdV equation \eqref{eq:kdv}, and other choices of nonlinear flux and dispersion recover well known scalar PDE models including the modified KdV \cite{miura_kortewegvries_1968}, Kawahara \cite{kawahara_oscillatory_1972}, and Gardner \cite{miura_kortewegvries_1968} equations for example. In addition to the original Whitham equation \eqref{eq:whitham_quadNL}, our approach applies to other nonlocal models including the Benjamin-Ono equation \cite{benjamin_internal_1967,ono_algebraic_1975}, Fornberg-Whitham equation \cite{fornberg_numerical_1978}, and the intermediate long wave equation \cite{joseph_solitary_1977}. 

We derive and analyze the Whitham modulation equations for the full dispersion generalized Whitham equation \eqref{eq:whitham}. The Whitham modulation equations are a system of conservation laws that describe the small dispersion limit of nonlinear wavetrains. Equivalently, they describe the slow evolution of a nonlinear, periodic wavetrain's parameters.  Various methods exist to derive the Whitham modulation equations including averaged conservation laws \cite{whitham_non-linear_1965}, averaged Lagrangian \cite{whitham_general_1965}, averaged Hamiltonian \cite{benzoni-gavage_modulated_2021}, or a multiple scale procedure \cite{luke_perturbation_1966}. For the KdV equation \eqref{eq:kdv}, the Whitham equations were proven to describe the zero dispersion limit \cite{lax_small_1983-2,lax_small_1983,lax_small_1983-1,venakides_zero-dispersion_1985} for $L^2(\mathbb{R})$ data with the inverse scattering transform.

The mathematical structure of the Whitham modulation system provides useful information about the evolution of a nonlinear periodic wavetrain. A particular application we focus on in this manuscript is the modulational instability (MI) of a periodic wavetrain. A history is provided in \cite{zakharov_modulation_2009}; we highlight some key discoveries. Benjamin and Feir demonstrated the breakup of nonlinear surface water waves propagating over a relatively deep, flat bottom \cite{benjamin_disintegration_1967,benjamin_instability_1967}. Simultaneously, Whitham developed his modulation theory and noted that the modulation equations may become elliptic \cite{whitham_linear_1974}. Subsequent work by Lighthill \cite{lighthill_contributions_1965} made the connection between ellipticity of the modulation equations and MI for weakly nonlinear dispersive waves, where the stability of the nonlinear wavetrain relies on the weakly nonlinear correction, $\omega_2$, to the linear dispersion relation and the linear dispersion curvature $\Omega''(k)$. This Lighthill-Whitham criterion 
\begin{align}\label{eq:ligthhill-whitham}
  \omega_2 \Omega''(k) < 0,
\end{align}
for modulationally unstable waves applies directly when there is no
induced mean flow. The introduction of wave-mean coupling requires a
modification of the analysis in \cite{whitham_linear_1974}. In both
cases, the coefficient $\omega_2$ and its wave-mean coupled
generalization are problem dependent. One significant result of the
present work is the generalized Lighthill-Whitham criterion:
$n(\ubar,k)\Omega''(k) < 0$ that explicitly depends upon $f(u)$ and
$\Omega(k)$. Its contraposative is a necessary condition for the
hyperbolicity, hence the modulational stability of weakly nonlinear
periodic waves. More recent theoretical developments have used a
spectral analysis of the linearized operator about a periodic
traveling wave solution to prove that weak hyperbolicity of the
modulation equations is a necessary condition for the modulational
stability of nonlinear periodic wavetrains
\cite{bronski_modulational_2010,benzoni-gavage_slow_2014,johnson_modulational_2020}. Note
that weak hyperbolicity only requires real characteristic eigenvalues
with no assumptions on the corresponding eigenvectors.  For
Hamiltonian equations, a stronger statement---modulational
stability---is available when the modulation equations are strictly
hyperbolic.

This manuscript is organized as follows. In Sec. \ref{sec:mainresults}, we summarize our main results: the Whitham modulation equations for the generalized Whitham equation \eqref{eq:whitham} and a generalized criterion for MI. In Sec. \ref{sec:derivation}, we carry out the derivation of the modulation equations. The modulation equations consist of averages that can be interpreted in the Fourier domain via Plancherel's theorem. The modulation equations in the small amplitude limit are the later focus of Sec. \ref{sec:derivation}, where we demonstrate that the system of modulation equations in the weakly nonlinear regime can be written as an explicit system that gives the approximate evolution of the modulation variables. In this regime, we identify properties of the quasilinear system and classify its hyperbolicity/ellipticity and genuine nonlinearity. The hyperbolicity/ellipticity of the weakly nonlinear modulation equations yields an index that specifies the modulational instability of a finite amplitude periodic wave. Modulational instability results are then related to those obtained from the Nonlinear Schr\"{o}dinger (NLS) approximation for the modulated waves yielding additional information such as the maximum growth rate and wavenumber associated with the instability. In Sec.~\ref{sec:comparisons}, we apply the MI index to various physical systems that can be modeled by Eq. \eqref{eq:whitham}. Finally, in Sec. \ref{sec:conclusions} we conclude the manuscript and discuss future problems related to the present work.

\section{Main results}
\label{sec:mainresults}
We consider the generalized Whitham equation \eqref{eq:whitham} with the following assumptions 
{\assumption \label{asump1} $\Omega(k)$ is a smooth, real valued, odd function and $\Omega'(0) = 0$. }
{\assumption \label{asump3} $f(u)$ is smooth and $f^{\prime}(0) = 0 $.}
{\assumption \label{asump4} There exists a three parameter family of $2\pi$ periodic traveling wave solutions $\varphi(\theta;\mathbf{p})$, $\theta = kx - \omega t$ with wavenumber $k(\mathbf{p})$ and frequency $\omega(\mathbf{p})$ to Eq. \eqref{eq:whitham} with $\mathbf{p} \in U \subset \mathbb{R}^3$. }\\

Assumptions \ref{asump1}--\ref{asump4} are satisfied by many models including the (m)KdV, generalized KdV, Gardner, Kawahara, Benjamin-Ono, Benjamin, and Whitham equations to name a few. 
We note that by Assumption 1, the linear dispersion relations $\Omega(k)$ we consider exclude those with small wavenumber asymptotics of the form $\Omega(k) \sim k^{\alpha}$, $\alpha < 1$. The spatial coordinate may be chosen so that parts of assumptions \ref{asump1} and \ref{asump3} hold without loss of generality. If $f'(0) = a$ and  $\Omega'(0) = b$ with $a,b \in \mathbb{R}$, transforming to the coordinate frame $\widetilde{x} = x - (a+b)t$, $\widetilde{t} = t$ removes the effect of these linear terms.

A challenge to derive the Whitham modulation equations for the generalized Whitham equation \eqref{eq:whitham} using standard approaches is the analysis of the nonlocal, convolution term. We found it helpful to conduct a multiple scale analysis in the Fourier domain. We assume a slowly modulated periodic, traveling wave solution of Eq. \eqref{eq:whitham} of the form 
\begin{align}
\label{eq:introptw}
u(x,t) = \varphi(\theta,X,T) + \epsilon u_1(\theta,X,T) + \ldots , \quad 0 < \epsilon \ll 1
\end{align}
where $X = \epsilon x$, $T = \epsilon t$ are long spatial and temporal scales, respectively and $\theta$ is the fast phase defined by $\theta = \theta(X,T)$
\begin{align}\label{eq:gen_phase}
\begin{split}\theta_t &= - \omega(X,T), \\\theta_x &= k(X,T),\end{split}\end{align} 
where $\omega(X,T)$ is the slowly evolving wave frequency and $k(X,T)$ is the slowly modulated wavenumber. We impose the periodicity requirement $\varphi(\theta,X,T) = \varphi(\theta+2\pi,X,T)$, $u_j(\theta,X,T) = u_j(\theta+2\pi,X,T)$ for $j = 1,2,3\ldots$, and $\theta,X,T \in \mathbb{R}$ so that we may express the leading order solution in terms of its Fourier series in $\theta$
\begin{equation}
\label{eq:fourser}
    \varphi(\theta,X,T) = \sum_{n = -\infty}^{\infty} \varphi_n(X,T) e^{i n \theta} \text{,}
\end{equation}
where the slowly varying Fourier coefficients $\varphi_n(X,T)$ admit the Fourier transform pair in the slow spatial scale
\begin{equation}
\label{eq:fcoeff}
    \varphi_n(X,T) = \frac{1}{2\pi}\int_{-\infty}^{\infty} \widehat{\varphi}_n(Q,T) e^{iQX}\mathrm{d}Q \text{,}
\end{equation}
\begin{equation*}
    \widehat{\varphi}_n(Q,T) = \int_{-\infty}^{\infty} \varphi_n(X,T) e^{-iQX}\mathrm{d}X \text{.}
\end{equation*}
In so doing, we identify the multiscale structure of the spatial wavenumber in the Fourier domain as the two-scale Fourier representation for $\varphi$
\begin{equation}\label{eq:two_scale_fourier}
  \begin{split}
    \varphi(\theta,X,T) &= \frac{1}{2\pi} \sum_{n = -\infty}^\infty
    \int_{\mathbb{R}} \widehat{\varphi}_n(Q,T) e^{i(n\theta + QX)} \,
    \mathrm{d} Q , \\
    \widehat{\varphi}_n(Q,T) &= \frac{1}{2\pi} \int_{-\pi}^\pi
    \int_{\mathbb{R}} \varphi(\theta,X,T) e^{-i(n\theta +
      QX)}\,\mathrm{d}X\mathrm{d}\theta ,
  \end{split}
\end{equation}
where integers $n$ correspond to harmonics of the rapidly varying, locally periodic wave and $Q$ is the modulation wavenumber.  Comparing the two scale Fourier representation \eqref{eq:two_scale_fourier} with the definitions \eqref{eq:fourier_trans_def}, we can identify the multiscale expansion of the spatial wavenumber in the Fourier domain for this slowly modulated periodic wave to be
\begin{equation}
  \label{eq:fourier_wavenumber}
  q = k n + \epsilon Q,
\end{equation}
i.e., small deviations in wavenumber from each harmonic. The multiscale expansion of the convolution operator is described in the following lemma.
\begin{lemma}
  \label{sec:operator_lemma}
  Assume a modulated periodic traveling wave solution to Eq. \eqref{eq:whitham} with analytic dispersion relation $\Omega(q)$ of the form \eqref{eq:introptw} with Fourier series representation \eqref{eq:fourser}. Then, the multiscale expansion of the convolution operator has the form
  \begin{equation}
    \label{eq:21}
    \mathcal{K}*u_x \sim k \mathcal{K}*\varphi_\theta +\epsilon  \left(\frac{1}{2}\left(\mathcal{K}'*\varphi_X + (\mathcal{K}'*\varphi)_X\right)  + k\mathcal{K}*u_{1,\theta}\right),
  \end{equation}
  where the $n^{\mathrm{th}}$ Fourier series coefficients are the phase velocity $\mathcal{K}_n = \frac{\Omega(nk)}{nk}$ and group velocity $\mathcal{K}'_n = \Omega'(nk)$ evaluated at the $n^{\textrm{th}}$ harmonic.
\end{lemma}
The proof of Lemma \ref{sec:operator_lemma} can be found in Appendix \ref{app:MS_nonlocal}. We remark that this result has the intuitively appealing interpretation that the two-scale expansion of the convolution appearing in the generalized Whitham equation \eqref{eq:whitham} in the Fourier domain acts upon the locally periodic wave in $\theta$ by multiplication of the linear phase velocity while acting upon the modulation in $X$ via multiplication by the group velocity. Utilizing Lemma \ref{sec:operator_lemma} and Plancherel's theorem, we derive the Whitham modulation equations for the generalized Whitham equation \eqref{eq:whitham} in Sec. \ref{sec:derivation}. 

\begin{proposition}
\label{prop:whitham_eqns}
Consider the generalized Whitham equation \eqref{eq:whitham} satisfying Assumptions \ref{asump1}-\ref{asump4}. The Whitham modulation equations for modulated wave solutions $\varphi(\theta;\mathbf{p}(X,T))$ are
\begin{subequations}\label{eq:whitham_eqs}
\begin{align}
\left( \overline{\varphi} \right)_T + \left(\overline{f(\varphi)}\right)_X & = 0 \label{eq:whitham_mass_intro}\\
\left(\frac{1}{2}\overline{\varphi^2}\right)_T + \left(\overline{F(\varphi)}+ \frac{1}{2}\overline{\varphi \mathcal{K}'*\varphi} \right)_X & = 0,\label{eq:whitham_momen_intro} \\
k_T + \omega_X & = 0 \label{eq:whitham_waves_intro}.
\end{align}
\end{subequations}
The averaging operator $\overline{G[\varphi]}(X,T)$ is defined by 
\begin{align}
\label{eq:avgop}
  \overline{G[\varphi]}(X,T) = \frac{1}{2\pi} \int_{-\pi}^\pi
  G\left[\varphi(\theta,X,T)\right]\,\mathrm{d}\theta ,
\end{align}
where $G$ is a local function of $\varphi$ and its derivatives,
\begin{align}
F(\varphi) = \int_0^{\varphi} s f'(s) \ {\rm d}s,
\end{align}
and
\begin{align}
\overline{\varphi\mathcal{K}'*\varphi} = \sum\limits_n  \Omega'(nk)|\varphi_n(X,T)|^2 .
\end{align}
\end{proposition}
An oft used set of physical modulation variables are  $\mathbf{p}  =[\ubar,a^2,k]^{\rm T}$ where $\bar{u}$ is the wave mean, $a$ is the waveheight, and $k$ is the spatial wavenumber. The parameters are related to the periodic wave according to 
\begin{subequations}
\label{eq:waveparams}
\begin{align}
\bar{u} &= \frac{1}{2\pi}\int_0^{2\pi} \varphi(\theta) d \theta, \\
a &= \max\limits_{\theta \in [0,2\pi)} \varphi(\theta) - \min\limits_{\theta \in [0,2\pi)} \varphi(\theta),\\
k & = \frac{2\pi}{L},
\end{align}
\end{subequations}
where $L$ is the spatial period of the periodic traveling wave. Two corollaries to Proposition \ref{prop:whitham_eqns} in the weakly nonlinear regime are the following. 

\begin{corollary}\label{cor:wnl_whitham}The Whitham modulation equations \eqref{eq:whitham_eqs} for $0 < a \ll 1$ are the conservation laws
\begin{subequations}\label{eq:weakly_NL_whitham_eqs}
\begin{align}
    \overline{u}_T + \left(f(\overline{u}) + \frac{a^2}{16}f^{\prime\prime}(\overline{u}) \right)_X &= 0, \label{eq:modeq1stokes}\\
    \left(\frac{1}{2}\overline{u}^2 + \frac{a^2}{16} + \frac{a^4 A^2}{64} \right)_T + \left(M_2 \right)_{X} &= 0,    \label{eq:modeq2stokes}\\
    k_T + \left(f'(\ubar)k + \Omega(k) + a^2 \omega_2 \right)_X & = 0,\label{eq:mod3stokes}
\end{align}
\end{subequations}
where
\begin{align}
\begin{split}
    M_2 &= F(\overline{u}) + \left(\frac{a^2}{16} + \frac{a^4 A^2}{64} \right)(f^{\prime}(\overline{u}) + uf^{\prime\prime}(\overline{u})) \\
&\quad    + \frac{a^2}{16}\Omega^{\prime}(k) + \frac{a^4 A^2}{64}\Omega^{\prime}(2k) \text{,}
\end{split}
\end{align}
and
\begin{subequations}\label{eq:Aandom2def}
    \begin{align}
    A &= \frac{kf''(\ubar)}{4\Omega(k) - 2\Omega(2k)}, \label{eq:A_def_intro} \\
    \omega_2 &= \frac{k}{16}\left(\frac{k\left(f''(\ubar)\right)^2}{2\Omega(k)-\Omega(2k)} + \frac{1}{2}f'''(\ubar) \right). \label{eq:omega2_def_intro}
    \end{align} 
\end{subequations}
\end{corollary}

\begin{corollary}\label{cor:whitham_properties}The Whitham modulation equations outlined in Corollary \ref{cor:wnl_whitham} are strictly hyperbolic if $\Omega^{\prime}(k) \neq 0$ and
\begin{align}
\label{eq:hypercrit}
    n(\ubar,k)\Omega''(k) > 0 \text{,}
\end{align}
where
\begin{align}
\label{eq:n_def}
\begin{split}
n(\ubar,k) &= k\left( \frac{(f''(\ubar))^2}{ \Omega'(k)} + \frac{k (f''(\ubar))^2}{2 \Omega(k) - \Omega(2k)}+  \frac{1}{2}f'''(\ubar)  \right)\\
& =  \frac{k(f''(\ubar))^2}{\Omega'(k)} + 16\omega_2
\end{split}
\end{align}
Moreover, the strictly hyperbolic Whitham modulation equations \eqref{eq:weakly_NL_whitham_eqs} are genuinely nonlinear if  $f''(\ubar) \neq 0$. 
\\

Assumption 1 may be relaxed, so that $\Omega(k)$ behaves like a linear function near the origin, i.e. $\Omega''(0)$ need not be zero. The corresponding Whitham modulation equations are hyperbolic provided 
\begin{align}
\label{eq:hypercrit_general}
    \widetilde{n}(\ubar,k)\Omega''(k) > 0 \text{,}
\end{align}
where
\begin{align}
\label{eq:ntilde_def}
\begin{split}
\widetilde{n}(\ubar,k) &=  k\left( \frac{(f''(\ubar))^2}{ \Omega'(k) - \Omega'(0)} + \frac{k (f''(\ubar))^2}{2 \Omega(k) - \Omega(2k)}+  \frac{1}{2}f'''(\ubar)  \right) \text{.}
\end{split}
\end{align}

\end{corollary}
Corollary \ref{cor:whitham_properties} is among the primary results in this manuscript, and it can be interpreted as a generalization of the aforementioned Lighthill-Whitham criterion \eqref{eq:ligthhill-whitham} to general nonlinear flux and linear dispersion with wave-mean coupling. 


In the weakly nonlinear regime,  modulations of a carrier wave can also be described by the Nonlinear Schr\"{o}dinger equation. The NLS approximation provides additional information about the growth of unstable perturbations. These are summarized in the following. 
\begin{corollary}\label{cor:NLS_approximation}
The slowly varying, complex envelope, $A$, of periodic solutions in the weakly nonlinear limit is governed by the NLS equation 
\begin{align}\label{eq:NLS_approx_cor}
   \begin{split} i A_\tau - n(\ubar,k) |A|^2 A + \frac{\Omega''(k)}{2}A_{\xi\xi} = 0,\\
   \tau = \epsilon^2 t, \quad \xi = \epsilon(x - \omega'(k) t),
   \end{split}
\end{align}
where $n(\ubar,k)$ is defined in \eqref{eq:n_def} and $\epsilon = \frac{a}{4}$. Perturbations to the wave envelope of the form 
\begin{align}
    A = e^{in\tau}\left(1 + \alpha e^{i\kappa \xi + \gamma \tau}\right), \quad |\alpha| \ll 1
\end{align}
grow exponentially if $n(\ubar,k)\Omega''(k) < 0$. The maximal growth rate occurs when $\kappa = \kappa_{\max}$
\begin{align*}
    \kappa_{\max}^2 = -\frac{n(\ubar,k)}{\Omega''(k)},
\end{align*}
and the corresponding maximal growth rate is 
\begin{align}
    \gamma_{\max} = |n(\ubar,k)|. 
\end{align}
\end{corollary}
Properties of the NLS approximation \eqref{eq:NLS_approx_cor} are discussed in Sec.~\ref{sec:NLS} and details of the derivation are given in Appendix \ref{app:NLS}. 

\section{Proofs of Proposition \ref{prop:whitham_eqns} and Corollaries 1 and 2}
\label{sec:derivation}
\begin{figure}[h]
\begin{center}
\includegraphics[scale=0.3]{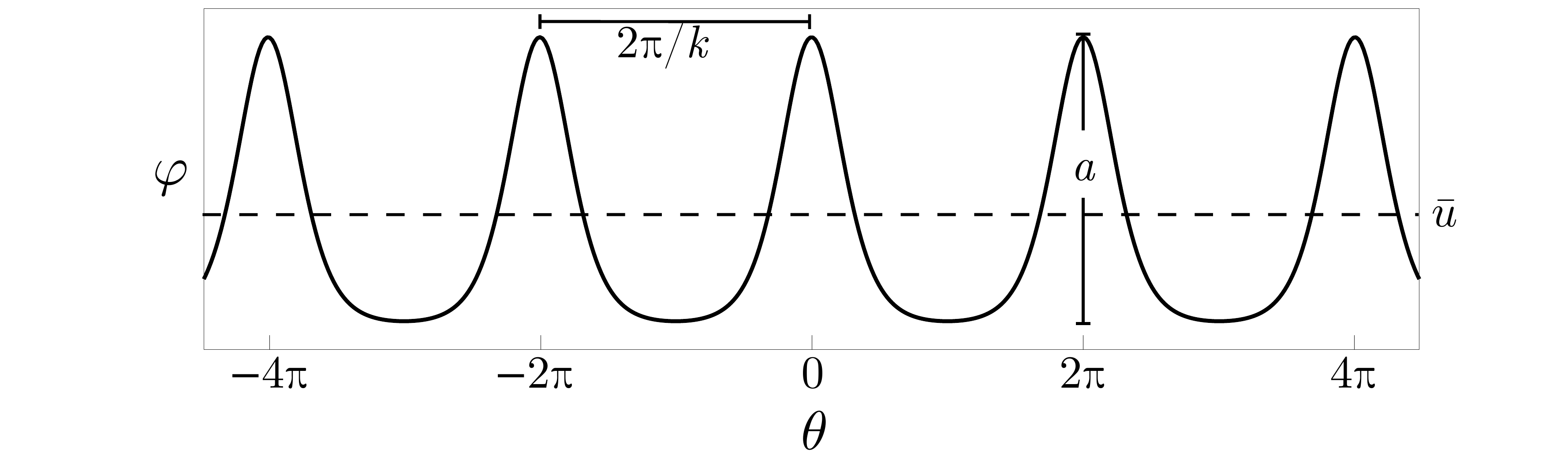}
\caption{Sketch of a nonlinear periodic wave with the physical parameters: wave mean $\ubar$, wave amplitude $a$, and wavenumber $k$. }
\label{fig:per_wave_sketch}
\end{center}
\end{figure}

Recall that the $2\pi$-periodic traveling waves $\varphi(\theta)$ are assumed to be characterized by three parameters: mean $\bar{u}$, amplitude $a$, and wavenumber $k$, as defined previously in Eq. \eqref{eq:waveparams}. Figure \ref{fig:per_wave_sketch} is a sketch of a nonlinear periodic wave with these parameters identified. 

The Whitham modulation equations \eqref{eq:whitham_eqs} are derived using a multiple scale procedure \cite{luke_perturbation_1966}. We seek a modulated $2\pi$-periodic traveling wave solution of the form of Eq. \eqref{eq:introptw} with phase satisfying Eq.   \eqref{eq:gen_phase}. Inserting the multiple scale expansion into Eq. \eqref{eq:whitham}, we collect and study   problems appearing at each increasing order in $\epsilon$. The leading order and first order equations are respectively 
\begin{align}
&\mathcal{O}(1):& -\omega \varphi_\theta + k f'(\varphi)\varphi_\theta + \mathcal{K}*(k\varphi_\theta) = 0 \label{eq:order_1} \\
&\mathcal{O}(\epsilon):& \mathcal{L} u_1 = -\varphi_{T} - f'(\varphi) \varphi_X - \frac{1}{2}\left[ \mathcal{K}'*\varphi_X +  \left(\mathcal{K}' *\varphi\right)_X \right] \label{eq:order_eps}
\end{align}
where the linear operator $\mathcal{L}$ is $$\mathcal{L} g =  -\omega g_\theta  +k \left(f'(\varphi)g \right)_\theta +  \mathcal{K}*(k g_\theta).$$
The leading order problem, Eq. \eqref{eq:order_1}, is the profile equation for the periodic traveling wave solution, $\varphi$, of Eq. \eqref{eq:whitham} that was assumed to exist.

The Whitham modulation equations are obtained by requiring a $2\pi$ periodic  solution to Eq. \eqref{eq:order_eps} to exist in $L^2\left([0,2\pi]\right)$. Equation \eqref{eq:order_eps} is solvable if the right hand side is orthogonal to the kernel of the adjoint operator $\mathcal{L}^{\dagger}$ 
\begin{equation}\label{eq:L_adjoint}
	\mathcal{L}^{\dagger}{g} =  \omega g_\theta - k f'(\varphi) g_\theta - \mathcal{K}*(k g_\theta) .
\end{equation}
The proof relies on integration by parts and application of the Plancherel theorem for the $L^2\left([0,2\pi]\right)$ inner product $\langle f,g\rangle$ of real functions $v$ and $w$ 
\begin{align*}
\langle v,w \rangle = \int_{0}^{2\pi} v(\theta) w(\theta) \ \mathrm{d}\theta .
\end{align*}
Then $\langle \mathcal{K}*v_\theta, w\rangle = 2\pi\sum_ni \Omega(kn) v_n w_n^*$ so that
\begin{align*}
    \langle \mathcal{L}v,w \rangle &= \int_{0}^{2\pi} \left(-\omega v_{\theta} w + k (f^{\prime}(\varphi) v)_{\theta}w + k(\mathcal{K}*v_{\theta}) w \right) \ \mathrm{d}\theta \\
    &= \omega \int_{0}^{2\pi} v w_{\theta} \ \mathrm{d}\theta - k\int_{0}^{2\pi} v f^{\prime}(\varphi) w_{\theta} \ \mathrm{d}\theta + 2\pi k \sum\limits_{n}i\Omega(nk)v_{n}w_{n}^{*} \\
    &= \langle v,\omega w_{\theta} \rangle + \langle v,-k f^{\prime}(\varphi)w_{\theta} \rangle -2\pi k \sum\limits_{n}v_{n} \left(i\Omega(nk)w_{n}\right)^{*} \\
    &= \langle v,\omega w_{\theta} \rangle + \langle v,-k f^{\prime}(\varphi)w_{\theta} \rangle + \langle v,-k \mathcal{K}*w_{\theta} \rangle = \langle v,\mathcal{L}^{\dagger}w \rangle \text{,}
\end{align*}
where $\cdot^*$ denotes complex conjugation.

By inspection, the kernel of $\mathcal{L}^{\dagger}$ \eqref{eq:L_adjoint} includes the linearly independent $2\pi$-periodic solutions $\varphi$ and $1$. Therefore, two necessary conditions for the existence of $2\pi$-periodic solutions to Eq. \eqref{eq:order_eps} are 
\begin{align}
	\label{eq:solve1}
\left\langle 1,\varphi_T + f^{\prime}(\varphi)\varphi_X + \frac{1}{2}\left[ \mathcal{K}'*\varphi_X +  \left(\mathcal{K}' *\varphi\right)_X \right]\right\rangle &= 0,\\
	\label{eq:solve2}
	 \left\langle \varphi , \varphi_T + f^{\prime}(\varphi) \varphi_X + \frac{1}{2}\left[ \mathcal{K}'*\varphi_X +  \left(\mathcal{K}' *\varphi\right)_X \right] \right\rangle  &= 0,
\end{align}
yielding the first two Whitham modulation equations. Averages involving the nonlocal operator can be computed directly by use of the Plancherel theorem. Adding the phase compatibility condition $\theta_{XT} = \theta_{TX}$, we arrive at the Whitham modulation equations \eqref{eq:whitham_eqs}.

The general approach here results in the Whitham modulation equations for any scalar equation that can be written as a generalized Whitham equation \eqref{eq:whitham}. Typically, the Whitham modulation equations are derived independently for each model equation. The approach presented here yields the Whitham modulation equations for an entire class of equations. The modulation equations \eqref{eq:whitham_eqs} exactly recover the modulation equations for canonical models such as the KdV equation with appropriate choices of $f$ and $\Omega$.  

We note that although we chose to use the method of multiple scales to derive the Whitham modulation equations \eqref{eq:whitham_eqs}, the same results can be obtained using the averaged conservation laws approach. The generalized Whitham equation \eqref{eq:whitham} admits the two conserved densities for decaying $u$ \cite{naumkin_nonlinear_1994}
\begin{subequations}
\begin{align}
  \label{eq:mass}
  \mathcal{P}_1 &= \int_{\mathbb{R}} u\, \mathrm{d}x, \\
  \label{eq:momen}
  \mathcal{P}_2 &= \int_{\mathbb{R}} u^2\, \mathrm{d}x.
\end{align}
\end{subequations}
 With these, we can obtain the first two Whitham modulation equations \eqref{eq:whitham_mass_intro} and \eqref{eq:whitham_momen_intro} by computing the averages
\begin{subequations}
\begin{align}
    \overline{u_t} + \overline{f(u)_x} + \overline{\mathcal{K}*u_x} &= 0, \\
    \overline{uu_t} + \overline{uf(u)_x} + \overline{u\mathcal{K}*u_x} &= 0,
\end{align}
\end{subequations}
using the averaging operation defined in Eq. \eqref{eq:avgop} and Lemma \ref{sec:operator_lemma} upon inserting the multiple scale expansion. Doing so results in Eqs. \eqref{eq:whitham_mass_intro} and \eqref{eq:whitham_momen_intro}, and the system is completed by adding the conservation of waves Eq. \eqref{eq:whitham_waves_intro}.

\subsection{Weakly nonlinear Whitham modulation equations}

In this section, we utilize a weakly nonlinear, Stokes wave approximation of periodic traveling wave solutions to the generalized Whitham equation \eqref{eq:whitham} and express the modulation equations \eqref{eq:whitham_eqs} explicitly in terms of parameters of the modulated periodic wave defined in Eq. \eqref{eq:waveparams}. The Stokes wave is computed upon seeking a periodic solution for small amplitude $0 < a \ll 1$. A standard perturbation calculation in Appendix \ref{app:stokes} demonstrates that the approximate periodic traveling wave profile up to $\mathcal{O}(a^2)$ is   
\begin{subequations}\label{eq:stokes}
\begin{align}
u &= \ubar + \frac{a}{2}\cos\theta + \frac{a^2}{4}A \cos2\theta + \mathcal{O}(a^3),\\
 \omega &= f'(\ubar)k + \Omega(k) + a^2 \omega_2 + \mathcal{O}(a^3)\label{eq:stokes_freq}
\end{align}
\end{subequations}
where $A$ and $\omega_2$ are given in equation \eqref{eq:Aandom2def}. The weakly nonlinear modulation equations are then determined by inserting the expansions \eqref{eq:stokes} into the modulation equations \eqref{eq:whitham_eqs}. Upon doing so, we obtain the Whitham modulation equations \eqref{eq:weakly_NL_whitham_eqs}. Up to this point, we retained all terms to $\mathcal{O}(a^2)$ when the modulation equations are cast in the quasilinear form
\begin{subequations}\label{eq:quasilinear_system_all}
\begin{align}\label{eq:quasilinear_form}
 &\hspace{4cm} \mathbf{p}_T + \mathbf{B} \mathbf{p}_X = 0, \\ \nonumber\\ 
\label{eq:Bmatrix}
    \mathbf{B} &=
    \begin{bmatrix} 
    f^{\prime}\left(\overline{u}\right) + \frac{1}{16}f^{\prime\prime\prime}\left(\overline{u}\right) a^2 &\frac{1}{16}f^{\prime\prime}\left(\overline{u}\right)& 0 \\
    2a^2 f''(\ubar) & b_{22} & a^2 \Omega''(k) \\
    f''(\ubar) k + a^2 \omega_{2,\overline{u}} & \omega_2 & f'(\ubar) + \Omega'(k)  + a^2 \omega_{2,k}
    \end{bmatrix}
    \text{,}\\
    b_{22} &= f^{\prime}(\overline{u}) + \Omega^{\prime}(k) + \frac{A^2}{2}(-\Omega^{\prime}(k) + \Omega^{\prime}(2k) + \ubar f^{\prime\prime}(\ubar))a^2,
\end{align}
\end{subequations}
where $\mathbf{p} = \left[ \ubar, a^2, k \right]^{\rm T}$.

\subsection{Mathematical properties of the weakly nonlinear Whitham modulation equations}\label{sec:properties}
The quasilinear, weakly nonlinear Whitham modulation equations
\eqref{eq:quasilinear_system_all} are in a form that is amenable to
further analysis, which has implications for the evolution of weakly
nonlinear wavetrains. Notable features of the modulation equations
\eqref{eq:quasilinear_system_all} we will identify here are their
hyperbolicity and genuine nonlinearity. Both of these properties rely
on the eigensystem for the matrix defined in
Eq. \eqref{eq:Bmatrix}. Generically, $\mathbf{B}$ possesses three
eigenpairs $\{\lambda_i,\mathbf{v}_i\}_{i = 1}^3$ where $\mathbf{v}_i$
is the right eigenvector in $\mathbb{R}^3$ with corresponding
eigenvalue $\lambda_i$. The following definitions are due to Lax
\cite{lax_hyperbolic_1973} (see also
Ref. \cite{dafermos_hyperbolic_2016}). The system
(\ref{eq:quasilinear_form}) is called \emph{hyperbolic} if its $3$
eigenvalues are real with linearly independent right eigenvectors. If
additionally, the 3 eigenvalues are distinct, then the system is
\emph{strictly hyperbolic}. If all eigenvalues are real but the
eigenvectors are linearly dependent, the system is sometimes referred
to as \emph{weakly hyperbolic} \cite{smoller_shock_1994}.  The
characteristic families of a hyperbolic system are then said to be
\emph{genuinely nonlinear} if
\begin{equation}
\label{eq:mugnl}
    \mu_j = \nabla \lambda_j \cdot \mathbf{v}_j \neq 0 \text{,} \ \ j = 1,2,3 \text{,}
\end{equation}
where the gradient $\nabla \lambda_j$ is taken with respect to the parameters $\ubar$, $a^2$, and $k$. If the system is not (weakly) hyperbolic, it is referred to as \emph{mixed elliptic} because there are necessarily two complex conjugate eigenvalues and one real eigenvalue.

We compute the approximate eigenvalues of the matrix \eqref{eq:Bmatrix} in a small amplitude expansion and obtain
\begin{subequations}
\label{eq:evals1}
\begin{align}
    \label{eq:lamb1}
    \lambda_1 &= f^{\prime}(\overline{u}) +         \frac{1}{16}\left(f^{\prime\prime\prime}(\overline{u}) + \frac{\left(f^{\prime\prime}(\overline{u})\right)^2(k\Omega^{\prime\prime}(k)-2\Omega^{\prime}(k))}{(\Omega^{\prime}(k))^2} \right)a^2 + \mathcal{O}(a^3), \\
    \label{eq:lamb2}
    \lambda_{2,3} &= f^{\prime}(\overline{u}) + \Omega^{\prime}(k) \pm \frac{a}{4}\sqrt{n(\ubar,k)\Omega^{\prime\prime}(k)} + \Lambda(\ubar,k) a^2 + \mathcal{O}(a^3),
\end{align}
\end{subequations}
where $n(\ubar,k)$ is defined in Eq. \eqref{eq:n_def}. The $\mathcal{O}(a^2)$ correction, $\Lambda(\ubar,k)$, to the eigenvalues corresponding to the split group velocity, $\lambda_2$ and $\lambda_3$, is real valued. Consequently, the precise value of this correction does not affect the mathematical structure of the quasilinear system \eqref{eq:quasilinear_form} unless higher order terms are retained. Therefore, we omit the explicit form of $\Lambda$. 

The asymptotic approximation of the eigenvalues \eqref{eq:evals1} is valid provided that $\Omega'(k) \neq 0$. This case will be addressed independently later in this section. We approximate the corresponding right eigenvectors as  
\begin{equation}
\label{eq:evec1}
    \mathbf{v}_1 = 
    \begin{bmatrix}
    -\frac{\Omega^{\prime}(k)}{k} \\
    0 \\
    f^{\prime\prime}(\overline{u})
    \end{bmatrix}
    + 
    \begin{bmatrix}
    v_{11}(\ubar,k) \\
    v_{12}(\ubar,k) \\
    0
    \end{bmatrix}a^2 + \mathcal{O}(a^3) \text{,}
\end{equation}
where
\begin{equation}
    v_{11}(\ubar,k) = \frac{v_{11,*}(\ubar,k)}{16 f^{\prime\prime}(\overline{u}) (k \Omega^{\prime}(k))^2} \text{,}
\end{equation}
\begin{equation}
    v_{12}(\ubar,k) = \frac{f^{\prime\prime}(\overline{u})(2\Omega^{\prime}(k) - k\Omega^{\prime\prime}(k))}{k\Omega^{\prime}(k)} \text{,}
\end{equation}
with
\begin{multline}
    v_{11,*}(\ubar,k) = f^{\prime\prime}(\overline{u})^3 k \left(-2\Omega^{\prime}(k) + k\Omega^{\prime\prime}(k) \right)\\ 
    + \Omega^{\prime}(k)^2\left(f^{\prime\prime\prime}(\overline{u})f^{\prime\prime}(\overline{u})k - 32f^{\prime\prime}(\overline{u})\omega_2(\ubar,k) + 16(-k f^{\prime\prime}(\overline{u})\omega_{2,k}(\ubar,k) + \omega_{2,\overline{u}}(\ubar,k)\Omega^{\prime}(k)) \right) \\
    +  16k\omega_2(\ubar,k)\Omega^{\prime}(k)\Omega^{\prime\prime}(k) .
\end{multline}
The remaining eigenvectors are
\begin{align}
    \label{eq:evec2}
    \mathbf{v}_{2,3} &= 
    \begin{bmatrix}
    0 \\
    0 \\
    1
    \end{bmatrix}
    \pm \sqrt{\frac{\Omega^{\prime\prime}(k)}{\Omega^{\prime}(k)^2 n(\ubar,k)}} \begin{bmatrix}
    \frac{1}{4}f^{\prime\prime}(\overline{u}) \\
    4\Omega^{\prime}(k) \\
    0
    \end{bmatrix}a + \mathbf{V}(\ubar,k) a^2 + \mathcal{O}(a^3),
\end{align}
where the $\mathcal{O}(a^2)$ correction, $\mathbf{V}(\ubar,k)$, to the right eigenvectors corresponding to the split group velocity, $\lambda_2$ and $\lambda_3$ is real valued. 

We observe that the quasilinear system \eqref{eq:quasilinear_form} is strictly hyperbolic if $n(\ubar,k)\Omega''(k) > 0$. 
If instead,
\begin{align}
\label{eq:instability_criterion}
    n(\ubar,k)\Omega^{\prime\prime}(k) < 0 \text{,}
\end{align}
then the eigenvalues $\lambda_2$ and $\lambda_3$ given in Eq. \eqref{eq:evals1} are complex conjugates. In this case, the weakly nonlinear Whitham modulation equations \eqref{eq:weakly_NL_whitham_eqs} are mixed elliptic. The level curves in the $\ubar$-$k$ plane across which $n(\ubar,k)\Omega''(k)$ changes sign correspond to certain physical mechanisms
\begin{itemize}
    \item $\Omega''(k) = 0$: An extremum of the group velocity 
    \item $\Omega'(k) = 0$, $k \neq 0$: short-long wave resonance due to coincident group velocity $\omega_k = f'(\ubar) + \Omega'(k)$ and long wavelength phase velocity $\lim\limits_{k\to 0}\frac{\omega(k)}{k} = f'(\ubar)$. 
    \item $2\Omega(k)=\Omega(2k)$: second harmonic resonance due to coincident phase velocities of the first and second harmonics. 
\end{itemize}
All of these mechanisms are independent of the mean $\ubar$. The remaining possibility is the level curve 
\begin{align}
    f''(\ubar)^2\left(2\Omega(k) - \Omega(2k) +k \Omega'(k)\right) + \frac{1}{2}f'''(\ubar)\Omega'(k) \left(2\Omega(k)-\Omega(2k)\right) = 0 \text{.}
\end{align}

A calculation shows that if the modulation equations \eqref{eq:weakly_NL_whitham_eqs} are strictly hyperbolic, all characteristic families are genuinely nonlinear if
\begin{align}
    f^{\prime\prime}(\overline{u}) \neq 0 \text{,}
\end{align}
which is the condition that appears in Corollary \ref{cor:whitham_properties}.  

At the short-long wave resonance corresponding to $\Omega'(k) =0$, $\mathbf{B}$ has 1 eigenvalue with algebraic multiplicity 3 at leading order. The asymptotic approximation of the eigensystem is correspondingly modified and we compute the distinct eigenvalues and corresponding right eigenvectors
\begin{equation}\label{eq:degenerate_evals}
    \lambda_j = f^{\prime}(\overline{u}) + \lvert k f^{\prime\prime}(k)^2 \Omega^{\prime\prime}(k)\rvert^{1/3} e^{2j\pi i/3} a^{2/3} + \mathcal{O}(a^{4/3}) \text{,} \ \ 
\end{equation}
\begin{equation}
    \mathbf{v}_j = \begin{bmatrix}
    0 \\
    0 \\
    1
    \end{bmatrix}
    + \begin{bmatrix}
    \lvert k f^{\prime\prime}(k)^2 \Omega^{\prime\prime}(k)\rvert^{1/3} e^{2j\pi i/3} \\
    0 \\
    0
    \end{bmatrix}a^{2/3} + \mathcal{O}(a^{4/3}) \text{,} \ \
\end{equation}
where $j = 0,1,2$. Two eigenvalues defined in \eqref{eq:degenerate_evals} are necessarily complex so the modulation system \eqref{eq:quasilinear_form} is mixed elliptic at the short-long wave resonance condition $\Omega'(k) = 0$. When $\Omega^{\prime\prime}(k) = 0$, the modulation system is weakly hyperbolic at $\mathcal{O}(a^2)$. A higher order analysis is required in order to determine the equation's type. If $2 \Omega(k) = \Omega(2k)$, further analysis is required to determine the system type because the Stokes expansion \eqref{eq:stokes} requires modification so that both the first and second harmonic modes appear at $\mathcal{O}(a)$. 
%

\section{Applications to modulational instability}
\label{sec:comparisons}
In this section, we discuss the modulational stability/instability of
periodic wavetrains that can be inferred from the Whitham modulation
equations. As noted earlier, weak hyperbolicity of the modulation
equations is necessary for the modulational stability of the
underlying periodic wave. In the weakly nonlinear regime, additional
details of the instability are determined by studying the Nonlinear
Schr\"{o}dinger (NLS) equation that describes the slowly evolving
nonlinear wave envelope (recall Corollary
\ref{cor:NLS_approximation}). In this section, we demonstrate that the
NLS approximation to the generalized Whitham equation
\eqref{eq:whitham} is focusing--which implies the periodic wave is
modulationally unstable--precisely when the weakly nonlinear Whitham
modulation equations \eqref{eq:weakly_NL_whitham_eqs} are mixed
elliptic.

\subsection{Nonlinear Schr\"{o}dinger equation approximation} 
\label{sec:NLS}
The NLS equation may be derived from Eq. \eqref{eq:whitham} upon seeking a slowly varying complex wave envelope of the form \cite{zakharov_multi-scale_1986,ablowitz_nonlinear_2011}
\begin{align}
u = \ubar + \epsilon \left[A e^{i\theta} + \cc \right] + \epsilon^2 \left[\frac{f''(\ubar)}{\Omega'(k)} |A|^2  + \frac{kf''(\ubar)}{2\Omega(k) - \Omega(2k)} A^2 e^{2i\theta} + \cc \right] + \mathcal{O}(\epsilon^3),
\end{align}  
where $0 < \epsilon = \frac{a}{4}\ll 1$ is a small amplitude scale. We introduce the scaled coordinate system 
\begin{align*}
 \xi =\epsilon \left(x - \omega'(k) t \right), \quad \tau = \epsilon^2 t,\quad \omega(k) = f'(\ubar) k + \Omega(k)
\end{align*}
and find that the complex wave envelope, $A(\xi,\tau)$, is described to leading order by the cubic NLS equation
\begin{align}\label{eq:NLS_equation}
i A_\tau -n(\ubar,k) |A|^2 A + \frac{\Omega''(k)}{2}A_{\xi\xi} &  = 0 \
\end{align}
where $n$ is defined exactly as in Eq. \eqref{eq:n_def}, which is provided again for reference
\begin{align}\tag{\ref{eq:n_def}}
n(\ubar,k) = k\left( \frac{(f''(\ubar))^2}{ \Omega'(k)} + \frac{k (f''(\ubar))^2}{2 \Omega(k) - \Omega(2k)}+  \frac{1}{2}f'''(\ubar)  \right).
\end{align}
The NLS equation \eqref{eq:NLS_equation} is defocusing if $n(\ubar,k)
\Omega''(k) > 0$ and focusing when $n(\ubar,k) \Omega''(k) < 0$. The
competition between nonlinearity and dispersion leads to MI when the
NLS equation \eqref{eq:NLS_equation} is focusing. As expected, the NLS
approximation to Eq. \eqref{eq:whitham} is focusing precisely when the
weakly nonlinear Whitham modulation equations
\eqref{eq:weakly_NL_whitham_eqs} are mixed elliptic, as defined by the
criterion given in Eq.~\eqref{eq:instability_criterion}. When
$n(\ubar,k) \Omega''(k) < 0$, the focusing NLS equation predicts the
growth rate of small amplitude perturbations to a steady periodic
traveling wave. The growth rate is identified by seeking a solution of
Eq.~\eqref{eq:NLS_equation} of the form
\begin{align}
    A = e^{i n(\ubar,k) \tau}\left(1 + \alpha e^{i\kappa \xi + \gamma \tau}  \right),
\end{align}
where $\alpha$ is the small, complex perturbation amplitude. We linearize about the stationary solution to find the relation between $\kappa$ and $\gamma$. Instability occurs when $\mathrm{Re} \ \gamma > 0$, where 
\begin{align}
 \gamma = \kappa \sqrt{-n(\ubar,k) \Omega''(k) - \frac{\left(\Omega''(k)\right)^2 \kappa^2}{4} },
\end{align}
which is real for perturbation wavenumbers $0 < \kappa^2 < \frac{-4n(\ubar,k)}{\Omega''(k)}$. The maximal growth rate of the instability occurs when $\kappa = \kappa_{\rm max}$
\begin{align}
    \kappa_{\rm max}^2 = -\frac{n(\ubar,k)}{\Omega''(k)} 
\end{align}
where the corresponding maximal growth rate is given by 
\begin{align}
    \gamma_{\mathrm{max}} = |n(\ubar,k)|.  
\end{align}

Note that the NLS equation \eqref{eq:NLS_equation} is valid when $n(\ubar,k)\Omega''(k) \neq 0$. At an inflection point of the dispersion, a higher order NLS approximation is required where an alternative criteria for MI can be obtained \cite{amiranashvili_extended_2019}. 

\subsection{Modulational instability of finite amplitude wavetrains in physical systems}
The remainder of this section is dedicated to studying the modulational instability of periodic wavetrains in several concrete physical systems. We will use the index $\tilde{n}(\ubar,k) \Omega''(k)$, where $\tilde{n}$ is defined in Eq.~\eqref{eq:ntilde_def} to determine curves in the $\ubar$-$k$ plane where wavetrains transition from stable to unstable. We specifically choose this quantity so that we do not need to subtract the long wave dispersion from $\Omega(k)$, i.e. $\Omega'(0)$ is not necessarily zero.  The only inputs for this analysis are the nonlinear flux and linear dispersion for the generalized Whitham equation \eqref{eq:whitham}. Table \ref{tab:applications} summarizes these nonlinear fluxes and dispersion relations for three applications. Relevant parameters will be defined in the subsequent sections when we present the stability results. We note that for these applications, we intentionally consider nonlinear flux functions $f(u)$ of a higher order than quadratic to demonstrate the improved accuracy and utility of the modulational instability index $\tilde{n}(\ubar,k) \Omega''(k)$.

\begin{table}[H]
\begin{center}
\begin{tabular}{p{3.75cm}cc}
\hline 
Application                 & $f(u)$ & $\Omega(k)$           \\ \hline\\
Gravity-capillary\newline water waves  &  $\frac{3 u^2}{4}-\frac{u^3}{8}$    & $\sqrt{(k + Bk^3) \tanh(k)}$                        \\[24pt]
Hydroelastic water \newline waves                            & $\frac{3 u^2}{4}-\frac{u^3}{8}$       & $\sqrt{(k + D k^5)\tanh{k}}$\\[24pt]
Internal waves & $ \frac{\alpha}{2} u^2 + \frac{\alpha_1}{3}u^3$ & $ \sqrt{\frac{k(1-\tilde{\rho})}{\tilde{\rho}\coth(k) + \coth(k \tilde{h}^{-1})}}$ 
\end{tabular}
\end{center}
\caption{Nonlinear flux and linear dispersion relation in the generalized Whitham equation \eqref{eq:whitham} for various problems.}
\label{tab:applications}
\end{table}

\subsubsection{Gravity-capillary water waves}
\label{sec:grav_cap}
As mentioned at the beginning of this section, the study of modulational instability of periodic water waves propagating on a finite depth dates back to the work of Benjamin \cite{benjamin_instability_1967} and Whitham \cite{whitham_linear_1974}, who, using independent methods, discovered the same transition from modulational stability to instability as the product of the undisturbed fluid depth, $h$, to the periodic wave's wavenumber $k$ passes through a critical threshold. For surface gravity waves, the threshold is
\begin{align} \label{eq:BF_instab}
kh = \eta_{\rm cr},
\end{align}
where $\eta_{\rm cr} \approx 1.363$. For $kh < \eta_{\rm cr}$, weakly nonlinear water waves are modulationally stable. Otherwise, they are modulationally unstable. In the subsequent discussion, equations are cast in nondimensional form (see ref. \cite{johnson_modern_1997}) such that $h = 1$. 

For pure gravity waves and weak nonlinearity, the nonlinear flux and dispersion are (cf. Eq. \eqref{eq:whitham_quadNL})
\begin{align}\label{eq:whitham_flux_disp}
f(u) = \frac{3}{4}u^2, \quad \Omega(k) = \sqrt{k \tanh k}, 
\end{align} 
which yields the original Whitham equation with the standard nondimensionalization so that the undisturbed fluid depth is unity. Small, but finite amplitude periodic solutions to the Whitham equations were proven to be unstable as the nondimensional wavenumber passes through $k \approx 1.146 <\eta_{\rm cr}$ \cite{hur_modulational_2015-1,sanford_stability_2014}. The index $n(\ubar,k)\Omega''(k)$ exactly reproduces this result. 

\begin{figure}[h!]
\begin{center}
\includegraphics[scale=0.45]{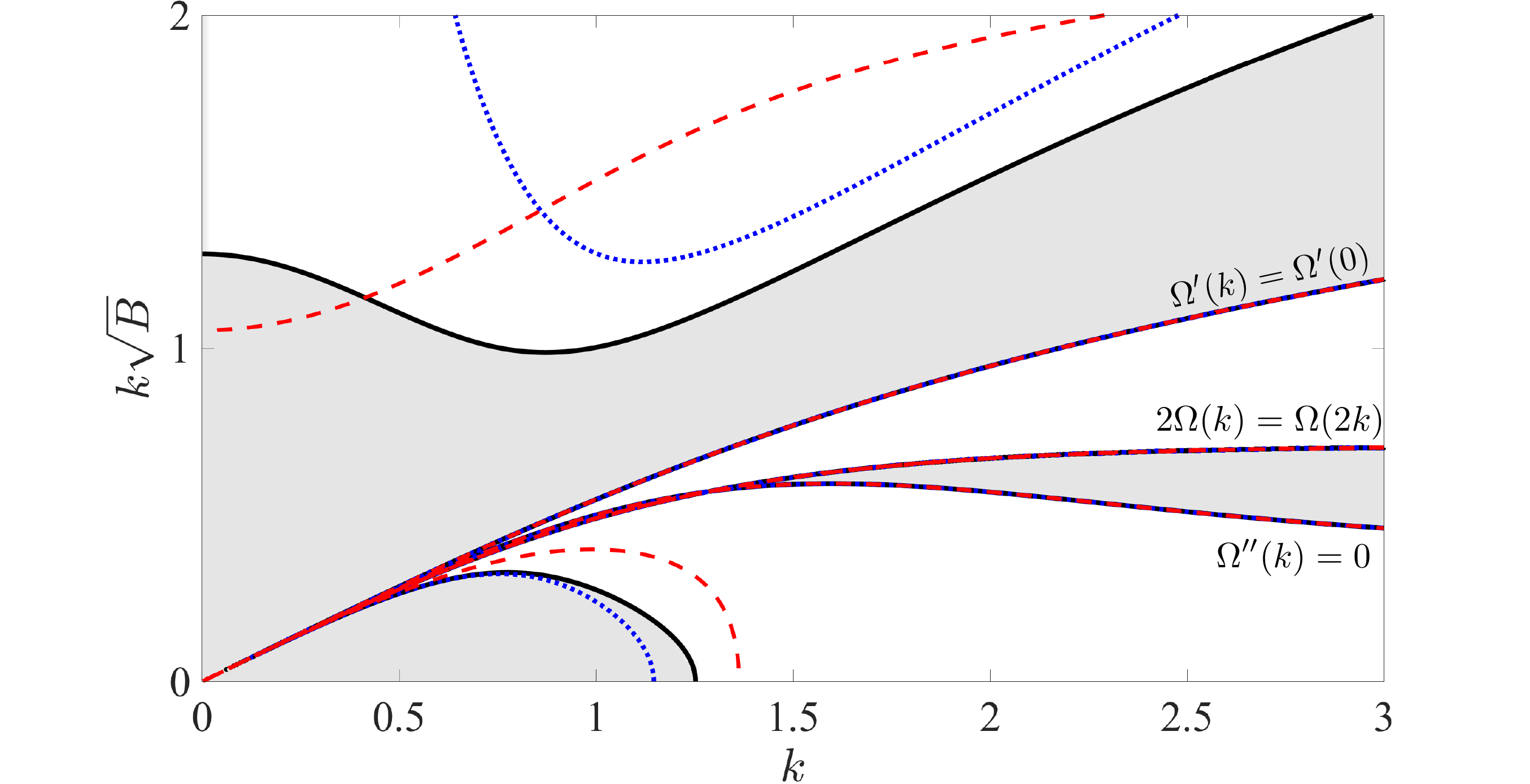}
\caption{Evaluation of the MI index $\tilde{n}(0,k)\Omega''(k)$ for gravity-capillary water waves modeled by a generalized Whitham equation with $f(u)$ and $\Omega(k)$ given in Table \ref{tab:applications}. Gray regions correspond to positive MI index waves (stable) and white regions coincide with negative MI index (unstable). Black, solid curves demarcate the border between the regions of stability/instability.  The dashed red curves indicate a change in the stability of a finite amplitude periodic traveling wave solution of the fully nonlinear Euler equations \cite{kawahara_nonlinear_1975}. The blue dotted curves represent boundaries separating stable and unstable periodic wave solutions of the Whitham equation with quadratic nonlinearity and water waves dispersion defined in Eq. \eqref{eq:whitham_flux_disp} (see ref. \cite{hur_modulational_2015-1}).}
\label{fig:grav_cap}
\end{center}
\end{figure}

Subsequent work incorporated surface tension into the Whitham equation, which appears solely in the linear dispersion relation. The simplest model of gravity capillary waves with full dispersion is the Whitham equation with quadratic nonlinear flux \eqref{eq:whitham_flux_disp} and the linear dispersion relation 
\begin{equation}\label{eq:grav_cap_disp}
    \Omega(k) = \sqrt{(k + Bk^3)\tanh k} ,
\end{equation}
where $B$, the Bond number, is a measure of the strength of the surface tension force in relation to the gravitational force. When $B = 0$, this model reduces to the Whitham equation for gravity water waves. The presence of surface tension results in a variety of novel phenomena, particularly near the critical value $B = 1/3$. Studies of the generalized Whitham equation \eqref{eq:whitham} with dispersion \eqref{eq:grav_cap_disp} and quadratic nonlinear flux \eqref{eq:whitham_flux_disp} established regions in the $(k,B)$ parameter space for which weakly nonlinear periodic traveling wave solutions are modulationally unstable \cite{hur_modulational_2015,carter_stability_2019}, see the dotted curves in Fig. \ref{fig:grav_cap}. 

We now incorporate a higher order nonlinear flux term into the
generalized Whitham equation \eqref{eq:whitham}. A convenient method
to obtain the flux was proposed by Whitham (see
\cite{whitham_linear_1974} p. 478). We retain nonlinear terms up to
third order in $u$ and consider the generalized Whitham equation
\eqref{eq:whitham} defined by the nonlinear flux and linear dispersion
given in Table \ref{tab:applications}. We compute the modulational
instability index $\tilde{n}(0,k)\Omega''(k)$ with $\ubar = 0$ and
compare the results to known results for the full water wave problem
\cite{kawahara_nonlinear_1975} and the case of quadratic flux in
Fig. \ref{fig:grav_cap}. Regions of $(k,B)$ parameter space where
periodic waves are modulationally stable appear in gray whereas regions that correspond to modulationally unstable waves are shown in white. The generalized Lighthill-Whitham criterion for modulational instability (solid curves) improves upon the quadratic flux results \cite{hur_modulational_2015} and is in good agreement with the stability analysis of solutions of the Euler equations with surface tension. The five red dashed curves correspond to a calculation of MI for the full Euler equations. The lowest solid curve intersects the $k$-axis where the growth rate becomes real (unstable), which approximates the cutoff for the Benjamin-Feir instability \eqref{eq:BF_instab}. As the nondimensional wavenumber passes through approximately $k \approx 1.252$, periodic waves in the absence of surface tension become modulationally unstable. The incorporation of cubic flux reduces the error in the critical wavenumber-depth ratio \eqref{eq:BF_instab} by 50$\%$ when compared to the prediction for quadratic flux ($k \approx 1.146$ \cite{hur_modulational_2015,sanford_stability_2014}). The MI index only includes derivatives of $f(\ubar)$ up to third order evaluated at $\ubar = 0$, so agreement with the stability analysis of the Euler equations will not be improved by including additional polynomial nonlinear terms.  
 The three middle curves demarcating the stable/unstable transition match exactly with those from the Euler equation since they depend only on the linear dispersion relation \cite{hur_modulational_2015}. This dependence is specified in Fig. \ref{fig:grav_cap}.

\subsubsection{Hydroelastic water waves}
\label{sec:flexural}
Hydroelastic, or flexural-gravity water waves refer to waves at the free surface of a body of water bounded by a deformable, elastic thin sheet. Physically, this may refer to the configuration where an inviscid, irrotational fluid is underneath a relatively thin ice sheet that is assumed to cause no friction with the water wave surface and modeled by an additional pressure term in the dynamic free surface boundary condition of the Euler equations \cite{plotnikov_modelling_2011}. Of particular interest in systems of this type is the influence of a moving, localized forcing. Experimental observation of moving vehicles over a frozen body of water demonstrate that typically a wide range of wavelengths are observed while the amplitude of the oscillations remains small \cite{takizawa_field_1988}. However, nonlinear effects, albeit weak, likely play a role in the wave evolution and various works demonstrate that nonlinearity has a marked effect on hydroelastic wave dynamics both for localized disturbances \cite{parau_nonlinear_2002,vanden-broeck_two-dimensional_2011} and periodic waves \cite{trichtchenko_stability_2019,xia_nonlinear_2002}. Recently, the Whitham equation with quadratic nonlinearity was investigated in this context to capture weak nonlinearity and full linear wave dispersion \cite{dinvay_whitham_2019} capturing qualitative features of solutions computed in the aforementioned studies. 

\begin{figure}[h!]
\begin{center}
\includegraphics[scale=0.45]{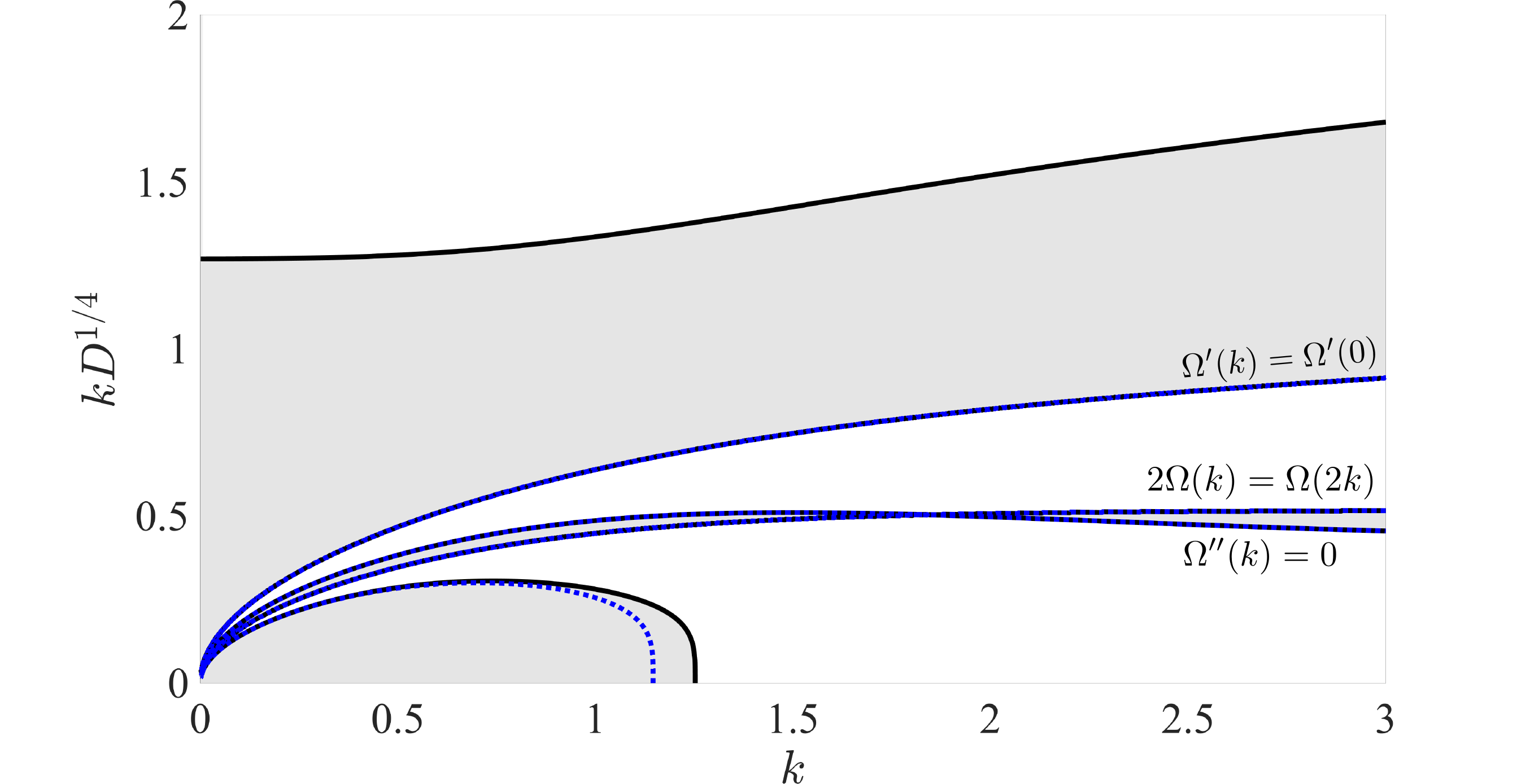}
\caption{Evaluation of the MI index $\tilde{n}(0,k)\Omega''(k)$ for flexural water waves modeled by a generalized Whitham equation  with $f(u)$ and $\Omega(k)$ given in Table \ref{tab:applications}. Gray regions correspond to positive MI index waves (stable) and white regions coincide with negative MI index (unstable). Black solid curves demarcate the border between the regions of stability/instability. The blue dotted curves are the boundaries separating stable and unstable periodic wave solutions of the Whitham equation for flexural water waves with quadratic nonlinearity $f(u) = \frac{3}{4} u^2$. }
\label{fig:flexural}
\end{center}
\end{figure}

The primary effect of the rigid ice sheet is to modify the linear dispersion, but not the nonlinear flux of free surface water waves \cite{xia_nonlinear_2002}. The generalized Whitham equation we propose here includes flux terms up to third order and linear dispersion defined in Table \ref{tab:applications}. The parameter $D > 0$ is a ratio of the flexural rigidity of the ice sheet and the density of the liquid underneath. In Fig. \ref{fig:flexural}, we plot regions where the index $n(0,k)\Omega''(k)$ indicates modulationally stable or unstable periodic solutions with zero mean.  In Fig. \ref{fig:flexural}, we have chosen axes to compare with Fig. \ref{fig:grav_cap}. 

\subsubsection{Internal waves}
\begin{figure}[h!]
\begin{center}
\includegraphics[scale=0.3]{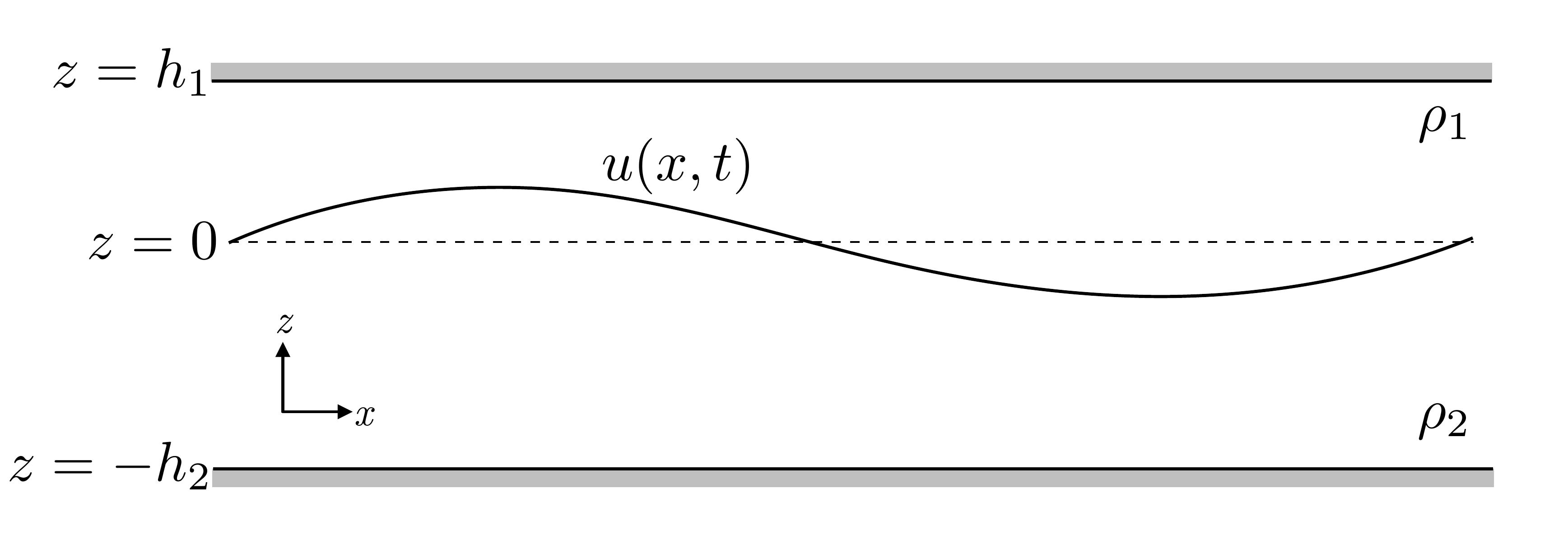}
\caption{Configuration for internal water waves at the free interface of two stratified fluids.}
\label{fig:internal_waves}
\end{center}
\end{figure}

We now consider the stability of internal waves for two immiscible fluid layers with differing densities. We identify the physical configuration where an incompressible fluid of density $\rho_1$ lies above a finite fluid layer of density $\rho_2$ with $\rho_2 > \rho_1$ to maintain a stable stratification. The undisturbed interface between the two fluids at the origin of the vertical coordinate $z = 0$ is bounded above by an impenetrable boundary at $z = +h_1$ and below by a similar boundary at $z = -h_2$. A sketch of the configuration considered here is shown in Fig. \ref{fig:internal_waves}

A long wave asymptotic expansion of the two fluid system when $h_2 \to \infty$ yields the Benjamin equation \cite{benjamin_new_1992}, which consists of the KdV third order dispersion term and a Hilbert transform term identical to that in the Benjamin-Ono equation. Other limits recover the KdV and intermediate-long wave equations \cite{choi_fully_1999}. Internal waves observed in the ocean are known to exhibit strong nonlinearity \cite{helfrich_long_2006}. Larger amplitude internal waves can be modeled by the Gardner equation with both quadratic and cubic nonlinear flux terms \cite{djordjevic_fission_1978}. Here, we retain up to cubic nonlinear terms and full linear dispersion, mirroring the previous two sections. In this physical system, we nondimensionalize coordinates according to
\begin{align}
    x\to h_1 \tilde{x}, \quad t \to \sqrt{\frac{h_1}{g}}\tilde{t},\quad u \to h_1 \tilde{u}, 
\end{align}
where $g$ is the gravitational acceleration. The nonlinear flux and linear dispersion of the generalized Whitham equation \eqref{eq:whitham} in nondimensional coordinates (upon dropping tildes from variables) are
\begin{subequations}\label{eq:internal_waves_coeffs}
\begin{align}
f(u) &= \frac{\alpha}{2} u^2 + \frac{\alpha_1}{3}u^3,\\ 
\Omega(k) &= \sqrt{\frac{k(1-\tilde{\rho})}{ \tilde{\rho}\coth k + \coth \frac{k}{\tilde{h}}}} ,
\end{align} 
where
\begin{align}
\alpha & = \frac{3}{2}c\frac{( \tilde{h}^2 - \tilde{\rho})}{(\tilde{\rho} + \tilde{h})} \label{eq:alpha_internal_waves},\\
\alpha_1 &= c\left[\frac{7}{8} \left(\frac{\tilde{\rho} -\tilde{h}^2}{\tilde{\rho} + \tilde{h}} \right)^2- \frac{\tilde{\rho}  + \tilde{h}^3}{\tilde{\rho}+  \tilde{h}}\right],\\
c & = \sqrt{ \frac{(1- \tilde{\rho})}{\tilde{\rho} +  \tilde{h}}},\\
\tilde{\rho}& = \frac{\rho_1}{\rho_2}, \\
\tilde{h} & = \frac{h_1}{h_2}. 
\end{align}
\end{subequations}
Note that the dispersion and nonlinear flux now depend on the two parameters $\tilde{\rho}$ and $\tilde{h}$. 

\begin{figure}[H]
\begin{center}
\includegraphics[scale=0.45]{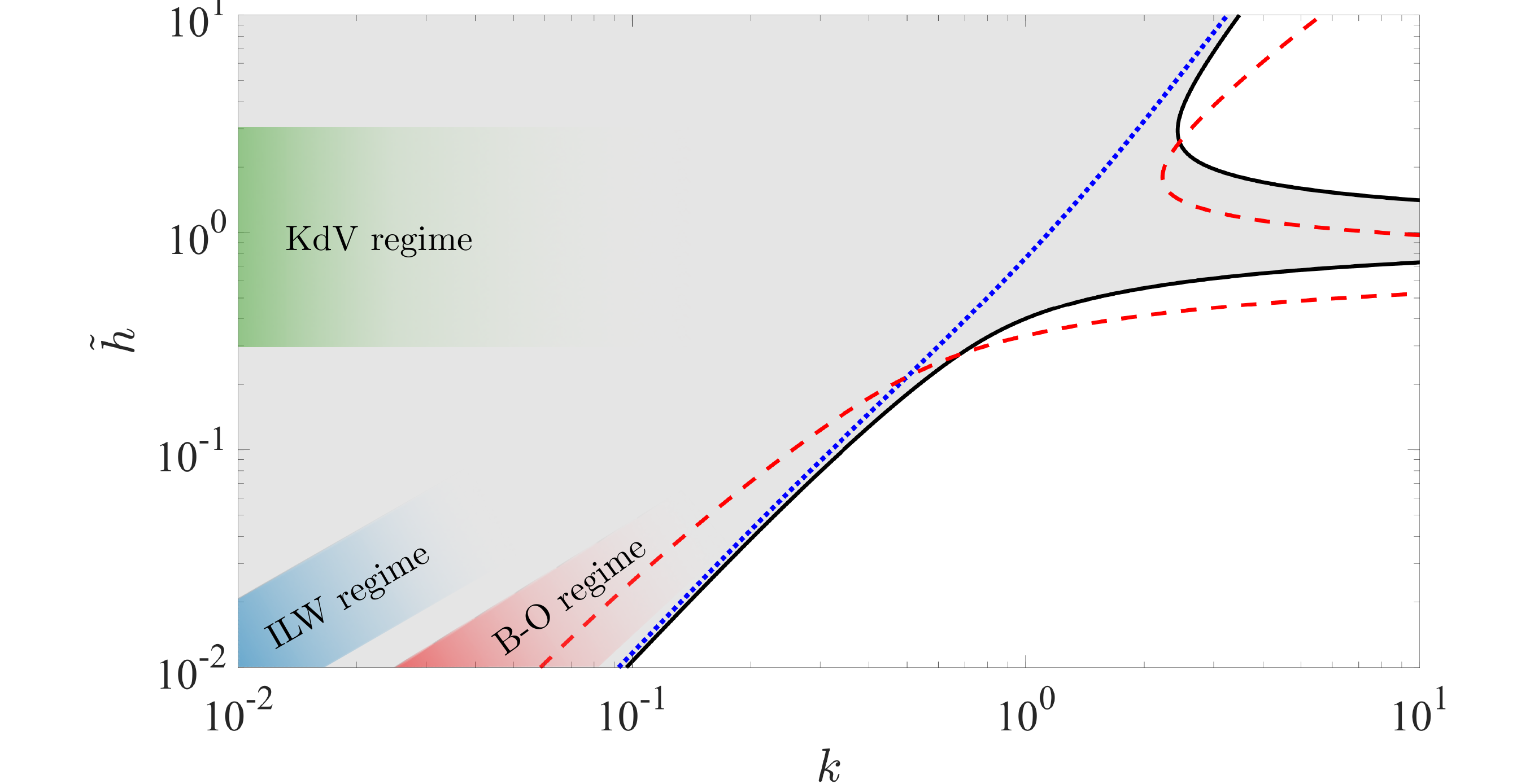}
\caption{Evaluation of the index $\tilde{n}(0,k)\Omega''(k)$ for waves at a two fluid interface modeled by a generalized Whitham equation with $f(u)$ and $\Omega(k)$ given in \eqref{eq:internal_waves_coeffs} for two fixed density ratios $\tilde{\rho}$. Gray regions correspond to positive MI index (stable) and white regions coincide with negative MI index (unstable). Black solid curves mark the border of the stable/unstable regions with density ratio $\tilde{\rho} = 0.99$.  The dashed red curves are the division of the stable/unstable regions for the density ratio $\tilde{\rho} =0.5$. The blue dotted curves are the boundaries separating stable and unstable periodic wave solutions of the Whitham equation for internal water waves with quadratic nonlinearity $f(u) =  \frac{\alpha}{2}u^2$ and density ratio $\tilde{\rho} = 0.99$. The colored shaded regions correspond to the scaling regimes of the oft used long-wave model equations where the periodic solutions are modulationally stable.  }
\label{fig:internal_waves_stability}
\end{center}
\end{figure}

We now compute examples of regions corresponding to stable/unstable waves for two fixed density ratios. We focus on the modulational stability of a range of wavelengths by plotting on a log-log scale for the two density ratios $\tilde{\rho} = 0.5$ and $\tilde{\rho} = 0.99$ shown in Fig. \ref{fig:internal_waves_stability}. Qualitatively, the curves separating the parameter regions corresponding to stable/unstable periodic waves are similar, though there are minor quantitative differences.  It is not surprising that long waves are modulationally stable since this is the region where the KdV, intermediate long wave (ILW), and Benjamin-Ono (B-O) equations are operable \cite{bona_asymptotic_2008}. The approximate parameter regimes where these models are applicable are shown in Fig. \ref{fig:internal_waves_stability}.  Note that the upper and lower bounds of the region of stable periodic waves approach $\tilde{h} = \sqrt{\tilde{\rho}}$ as $k \to \infty$ when cubic nonlinearity is included. This coincides with parameter values for which the coefficient of the quadratic nonlinear (cf. Eq. \eqref{eq:alpha_internal_waves}) term vanishes. The blue dotted curve in Fig. \ref{fig:internal_waves_stability} separates parameter regimes corresponding to stable/unstable periodic traveling wave solutions of the generalized Whitham equation with quadratic nonlinearity $f(u) = \frac{\alpha}{2}u^2$ and linear dispersion for internal waves with density ratio $\tilde{\rho} = 0.99$. 

\section{Discussion and conclusions}
\label{sec:conclusions}
In this manuscript, we derived the Whitham modulation equations for nonlinear, fully dispersive scalar model equations we called the generalized Whitham equation \eqref{eq:whitham}. The presence of the nonlocal dispersive term in this model motivated a Fourier based multiple scale approach to derive the modulation equations. The modulation equations are general but concrete, involving averages over the periodic traveling wave manifold. 

We then considered the Whitham modulation equations in a weakly nonlinear regime, where periodic traveling wave solutions to the generalized Whitham equation were obtained. In so doing, we determined the modulation equations in explicit, quasilinear form. From there, we derived conditions under which the system was strictly hyperbolic or mixed elliptic and genuinely nonlinear.

Finally, we considered the application of this theory to modulational instability. By deriving a criterion for hyperbolicity of the modulation equations in the weakly nonlinear regime, we obtained an explicit generalized Lighthill-Whitham criterion for modulational instability that depends explicitly on the nonlinear flux $f(u)$ and linear dispersion relation $\Omega(k)$. We applied this MI index to various geophysical fluid applications (gravity-capillary water waves, hydroelastic water waves, and internal waves), demonstrating the significance of higher order nonlinearity that is determinant of the stability boundaries.

Our results are applicable to many physical models by simply specifying the nonlinear flux $f(u)$ and linear dispersion relation $\Omega(k)$, subject to modest assumptions. Moreover, we derived the NLS equation in the weakly nonlinear regime from the generalized model Eq.~\eqref{eq:whitham}. The nonlinear coefficient $n(\ubar,k)$ is found by selecting the nonlinear flux $f(u)$ and linear dispersion relation $\Omega(k)$ in Eq.~\eqref{eq:n_def}, as opposed to full asymptotic re-derivation for different nonlinear fluxes. As such, the results given in this work are of particular interest to those with applications where full dispersion scalar models are used.

There are several potential directions for future research. While we analyzed the modulation equations in the weakly nonlinear regime, an analysis of the modulation equations in the fully nonlinear regime is of interest as well. More generally, the ease of applying the results of Corollary \ref{cor:whitham_properties} to general full dispersion scalar models means that this work could be extended to applications beyond those we discussed in Section \ref{sec:comparisons}. Explicit expressions for families of periodic solutions are not always available, so properties of the modulation equations can be determined numerically. These numerical computations have been carried out for the conduit equation \cite{maiden_modulations_2016} and the fifth order KdV equation \cite{sprenger_discontinuous_2020}.

The Whitham modulation equations \eqref{eq:whitham_eqs} can be used to study dynamical solutions of certain initial value problems. A common application of Whitham modulation theory is to study the structure of dispersive shock waves (DSWs), which typically consist of an expanding, modulated periodic wave with a harmonic wavetrain at one edge and an approximate solitary wave at the opposite edge. The Whitham modulation equations can be used to determine both macroscopic properties, e.g. the velocities of the disparate edges, and the DSW's interior structure \cite{el_dispersive_2016}. Novel dynamics in scalar models containing cubic nonlinear flux \cite{el_dispersive_2017-1} or higher order/nonlocal dispersion \cite{sprenger_shock_2017,el_dispersive_2018,hoefer_modulation_2019} have been observed. These include so-called Whitham shocks--discontinuous weak solutions of the modulation equations that correspond to zero dispersion limits of periodic heteroclinic to periodic traveling wave solutions \cite{sprenger_discontinuous_2020}. A promising research direction is the consideration of Whitham shocks for the generalized Whitham equation \eqref{eq:whitham}. The derivation of the general NLS equation with nonlinear coefficient $n(\ubar,k)$ given in Eq. \eqref{eq:n_def} can be used for a universal description of the DSW structure near the harmonic edge \cite{congy_nonlinear_2019}. The general NLS equation presented in this manuscript allows for this approach to be used for any model equation that can be cast in the form of Eq. \eqref{eq:whitham}. 


Finally, the Fourier based approach used to derive the Whitham modulation equations can be applied to study more general full dispersion equations. These may include models incorporating nonlinear dispersive terms (for instance of Camassa-Holm type \cite{lannes_water_2013}), Boussinesq systems \cite{hur_modulational_2019,aceves-sanchez_numerical_2013}, or full dispersion NLS models \cite{naumkin_nonlinear_1994}.

\section*{Acknowledgments}
This work was partially supported by NSF grants DMS-1816934  and DMS-1812445.

\providecommand{\bysame}{\leavevmode\hbox to3em{\hrulefill}\thinspace}
\providecommand{\MR}{\relax\ifhmode\unskip\space\fi MR }
\providecommand{\MRhref}[2]{%
  \href{http://www.ams.org/mathscinet-getitem?mr=#1}{#2}
}
\providecommand{\href}[2]{#2}

\appendix 

\section{Multiple scale expansion of the nonlocal convolution operator}
\label{app:MS_nonlocal}
The multiple scale expansion of the convolution operator in Sec.~\ref{sec:derivation} is required to derive the Whitham modulation equations \eqref{eq:whitham_eqs}. In this appendix we derive this expansion. For the following calculation, we assume the existence of a periodic traveling wave solution to Eq. \eqref{eq:whitham} where the dispersion relation $\Omega$ is an analytic function of its argument and $\Omega(0) = \Omega'(0) = 0$. 

The modulation equations are derived upon assuming the existence of a  periodic wave with the slowly varying ansatz
\begin{align}
  \label{eq:slowly_varying_per_wave}
  \begin{split}
  u(x,t) = \varphi(\theta,X,T) \equiv \varphi(\theta;\mathbf{p}(X,T))\\ \quad X
  = \epsilon x, \quad T = \epsilon t, \quad 0 < \epsilon \ll 1,
  \end{split}
\end{align}
in which the vector of parameters $\mathbf{p}$ varies on the slow
scales $X$ and $T$ while $\varphi$ remains $2\pi$-periodic in
$\theta$: $\varphi(\theta+2\pi,X,T) = \varphi(\theta,X,T)$, for all
$\theta, X \in \mathbb{R}$, $T > 0$.  This multiscale ansatz \eqref{eq:slowly_varying_per_wave} results in the usual differential operator expansions
\begin{equation}
  \label{eq:operator_expansion}
  \partial_x = k \partial_\theta + \epsilon \partial_X,
  \quad \partial_t = -\omega \partial_\theta + \epsilon \partial_T .
\end{equation}
Within this context, it is convenient to express the nonlocal convolution operator as the following pseudodifferential operator 
\begin{align}
\mathcal{K}*\varphi_x = i \Omega(-ik \partial_x)\varphi, 
\end{align}
We replace $\partial_x$ with the multiscale expansion and expand $\Omega$ about the origin to obtain
\begin{align}
\begin{split}\label{eq:A4}
i \Omega(-ik \partial_x)\varphi &\to i \Omega(-ik \partial_\theta - i \epsilon \partial_X)\varphi\\
& = i\left[ \frac{\Omega''(0)}{2}(-i k \partial_\theta - i \epsilon \partial_X)^2 +  \frac{\Omega'''(0)}{6}(-i k \partial_\theta - i \epsilon \partial_X)^3 \right. \\ 
& \quad + \left. \frac{\Omega^{(4)}(0)}{24}(-i k \partial_\theta - i \epsilon \partial_X)^4 + \ldots \right] \varphi.  
\end{split}
\end{align}
A calculation reveals that the $\mathcal{O}(1)$ and $\mathcal{O}(\epsilon)$ terms in expansion may be expressed by  
\begin{align}\label{eq:A5}
\begin{split}
(-i k \partial_\theta - i \epsilon \partial_X)^n\varphi & \sim (- i k \partial_\theta)^n \\ &  + \frac{\epsilon}{2} \left(n (- i k \partial_\theta)^{n-1} \varphi_X  + \partial_X \left((-i k\partial_\theta)^n \varphi \right)\right),
\end{split}
\end{align}
Replacing the polynomial terms in Eq. \eqref{eq:A4} with the expansions \eqref{eq:A5} and recognizing the series expansions of $\Omega$ and $\Omega'$, we determine 
\begin{align}\label{eq:A6}
i \Omega(-ik \partial_\theta - i \epsilon \partial_X)\varphi \sim i \Omega(-ik \partial_\theta) \varphi + \frac{\epsilon}{2} \left[\Omega'(-ik\partial_\theta) \varphi_X + \partial_X (\Omega'(-ik\partial_\theta) \varphi)\right]. 
\end{align}
The expansion \eqref{eq:A6} can be expressed in terms of convolution operators, and we therefore find that the multiple scales expansions of the nonlocal convolution term appearing in Eq. \eqref{eq:whitham} is 
\begin{align}\label{eq:A7}
\mathcal{K}*\varphi_x \sim k \mathcal{K}*(\varphi_\theta) + \frac{\epsilon}{2} \left(\mathcal{K}'*\varphi_X + (\mathcal{K}'*\varphi)_X\right), 
\end{align}
where $\mathcal{K}'*\varphi$ is defined by the Fourier multiplier 
\begin{align}
\widehat{\mathcal{K}'*\varphi} = \Omega'(q) \widehat{\varphi}. 
\end{align}
The result given in Eq. \eqref{eq:A7} is the expansion used in Lemma \ref{sec:operator_lemma}. 

\section{Stokes Expansion}
\label{app:stokes}
In this appendix, we compute an approximate periodic, traveling wave solution to Eq. \eqref{eq:order_1} via an asymptotic expansion in the wave amplitude, $a$. The expansion is of the form  
\begin{equation}
\label{eq:uexpand}
    \varphi =  \ubar + \frac{a}{2} \varphi_1 + a^2\varphi_2 + \ldots \text{,}
\end{equation}
\begin{equation}
\label{eq:omstokesexpand}
    \omega = \omega_0 + a \omega_1 + a^2 \omega_2 + \ldots \text{,}
\end{equation}
where $\overline{u}$ is the mean and $\omega_0$ is the linear dispersion relation 
\begin{align}
    \omega_0(\ubar,k) = f'(\ubar) k + \Omega(k). 
\end{align}
We insert the asymptotic expansions \eqref{eq:uexpand} and \eqref{eq:omstokesexpand} into \eqref{eq:order_1} and collect in powers of the amplitude. The leading order problem is trivially satisfied, and the first nontrivial problem appears at $\mathcal{O}(\epsilon)$ where we have
\begin{equation}
\label{eq:u1ode}
    -\omega_0 \varphi_1^{\prime} + k f^{\prime}(\overline{u}) \varphi_1^{\prime} + \mathcal{K}*(k \varphi_1^{\prime}) = 0,
\end{equation}
where primes denote derivatives with respect to $\theta$. The integro-differential equation can be solved with Fourier transforms. The $2\pi$ periodic solution at leading order is given by 
\begin{align*}
\varphi_1 = \cos{\theta} .
\end{align*}
We continue to the next order $\mathcal{O}(a^2)$ where we have the intego-differential equation for $\varphi_2$
\begin{equation}
\label{eq:ftoeps2}
    -\omega_0 \varphi_2^{\prime} + k f^{\prime}(\overline{u}) \varphi_2^{\prime} + \mathcal{K}*(k \varphi_2^{\prime}) = \omega_1 \varphi_1^{\prime} - kf^{\prime\prime}(\overline{u}) \varphi_1 \varphi_1^{\prime}. 
\end{equation}
The linear equation is solvable provided the forcing terms are orthogonal to the kernel of the linear operator that defines the integro-differential equation. Secular terms are removed so long as $\omega_1 = 0$. We solve Eq.~\eqref{eq:ftoeps2} with Fourier transforms and find
\begin{equation}
\varphi_2 =  \frac{k f^{\prime\prime}(\overline{u})}{4\Omega(k) - 2\Omega(2k)}\sin(2\theta)
\end{equation}
The $\mathcal{O}(a^2)$ correction to the frequency is found by ensuring no secular terms appear at $\mathcal{O}(\epsilon^3)$ 
\begin{align}
\label{eq:lu3}
\begin{split}
    &-\omega_0 \varphi_3^{\prime} + k f^{\prime}(\overline{u}) \varphi_3^{\prime} + \mathcal{K}*(k \varphi_3^{\prime}) \\& = \omega_2 \varphi_1^{\prime} - kf^{\prime\prime}(\overline{u}) \varphi_1 \varphi_2^{\prime} - kf^{\prime\prime}(\overline{u}) \varphi_2 \varphi_1^{\prime} -\frac{kf^{\prime\prime\prime}(\overline{u})}{2}\varphi_1^2 \varphi_1^{\prime}.
    \end{split}
\end{align}
Secular terms--coefficients of $\cos(\theta)$ terms--are removed upon setting
\begin{equation}
    \omega_2(k) = \frac{k}{4}\left(\frac{k\left(f^{\prime\prime}(\overline{u})\right)^2}{2\Omega(k)-\Omega(2k)} + \frac{f^{\prime\prime\prime}(\overline{u})}{2} \right) \text{.}
\end{equation}
Therefore, the approximation for the weakly nonlinear periodic solution is given by 
\begin{equation}
\label{eq:stokeswave}
    u = \overline{u} + \frac{a}{2}\cos{\theta} + \frac{k f^{\prime\prime}(\overline{u})}{16\Omega(k) - 8\Omega(2k)}\cos{2\theta} + \mathcal{O}(a^3)
\end{equation}
with frequency
\begin{align}
    \omega(k,\overline{u},a)&= kf^{\prime}(\overline{u}) + \Omega(k) + \frac{a^2k}{16}\left(\frac{k\left(f^{\prime\prime}(\overline{u})\right)^2}{2\Omega(k)-\Omega(2k)} + \frac{f^{\prime\prime\prime}(\overline{u})}{2} \right) + \mathcal{O}(a^3),
\end{align}
where $a \ll 1$ for the asymptotic series to be well ordered. Note that this result applies for any general scalar model equation with appropriate $f(u)$ and $\Omega(k)$. Note that the Stokes solution is invalid if $2\Omega(k) = \Omega(2k)$, which corresponds to a resonance due to coincident phase velocities at the first and second harmonic.

\section{Derivation of the NLS equation}
\label{app:NLS}
The Nonlinear Schr\"{o}dinger (NLS) equation describes the weakly nonlinear complex envelope of a carrier wave. The asymptotic derivation of the NLS equation assumed a solution in a small parameter $0 < \epsilon = \frac{a}{4} \ll 1$ of the form 
\begin{align*}
u = \ubar(\zeta,X,T) +  \epsilon u_1(\zeta,X,T) + \epsilon^2 u_2(\zeta,X,T) + \ldots,
\end{align*}
where $X$ and $T$ are the slow space and time scales introduced previously and $\zeta = kx - \left(f'(\ubar)k + \Omega(k)\right) t$. At $\mathcal{O}(\epsilon)$, we obtain the linear, homogeneous equation 
\begin{align}\label{eq:NLS_Mu1}
\mathcal{M}u_1 :=\left\{-\Omega(k) \partial_\zeta  + \mathcal{K}\partial_\zeta\right\}u_1 = 0 
\end{align}
which has the solution $u_1 = A(X,T) e^{i\zeta} + \cc + M(X,T)$, where $\cc$ denotes the complex conjugate and $M(X,T)$ is the slowly varying mean. At $\mathcal{O}(\epsilon^2)$ we have the nonhomogeneous equation 
\begin{align}\label{eq:NLS_Mu2}
\mathcal{M}(u_2) &=  -u_{1,T} - f'(\ubar) u_{1,X} - \Omega'(-i k \partial_\zeta) u_{1,X} - k f''(\ubar) u_1 u_{1,\zeta}, 
\end{align}
which contains no secular terms provided that 
\begin{align}
A_T + \omega'A_X + i k f''(\ubar)MA & = \epsilon G_1 + \epsilon^2 G_2 +\ldots \\
M_T & = \epsilon F_1 + \epsilon^2 F_2 + \ldots,
\end{align}
where we have introduced the higher order corrections $G_i$ and $F_i$, $i = 1,2,3\ldots$. We solve Eq. \eqref{eq:NLS_Mu2} and find 
\begin{align}\label{eq:NLS_u2}
u_2 =  \frac{k f''(\ubar) A^2}{2 \Omega(k) - \Omega(2k)}e^{2i\zeta} + \cc. 
\end{align}
We continue the asymptotic calculation to the next order in $\epsilon$. At this order, solvability with respect to the first harmonic ($e^{i\zeta}$) and constant forcing terms result respectively in 
\begin{align}
\begin{split}
&A_T + \omega'A_X + k f''(\ubar)MA \\& =\epsilon\left( -i\frac{(k f''(\ubar))^2}{2 \Omega(k) - \Omega(2k)}- i \frac{k}{2}f'''(\ubar) \right)|A|^2 A  + i \epsilon \frac{\Omega''(k)}{2}A_{XX},\label{eq:NLS_AMeq}
\end{split}\\
&M_T + f'(\ubar) M_X = - \epsilon f''(\ubar)\left(|A|^2\right)_X,\label{eq:NLS_Meq}
\end{align}
which shows that $M$ is $\mathcal{O}(\epsilon)$ in the asymptotic expansion. 
Upon introducing the variables
\begin{align*}
\xi = X - \omega'(k) T, \quad \tau = \epsilon T,
\end{align*}
we find that, to $\mathcal{O}(\epsilon)$, Eq.~\eqref{eq:NLS_Meq} can be solved to find that 
\begin{align}\label{eq:NLS_M}
M = \epsilon \frac{f''(\ubar)}{\Omega'(k)}|A|^2. 
\end{align} 
Replacing $M$ in Eq.~\eqref{eq:NLS_AMeq} with the expression \eqref{eq:NLS_M} we find that the nonlinear envelope, $A$, solves the NLS equation (cf. Eq.~\eqref{eq:NLS_equation})
\begin{align}
i A_\tau - n(\ubar,k) |A|^2 A + \frac{\Omega''(k)}{2}A_{\xi\xi} &  = 0 \
\end{align}
where 
\begin{align*}
n(\ubar,k) =  k\left( \frac{(f''(\ubar))^2}{ \Omega'(k)} + \frac{k (f''(\ubar))^2}{2 \Omega(k) - \Omega(2k)}+  \frac{1}{2}f'''(\ubar)  \right).
\end{align*}

\end{document}